\documentclass[10pt,letterpaper,twocolumn]{article} 

\usepackage{ol2}
\usepackage[draft]{hyperref}
\usepackage{amsmath}

\begin{document}

\twocolumn[ 

\title{Creation of arbitrary spectra with an electro-optic modulator}


\author{C. E. Rogers III,$^{1}$ J. L. Carini,$^{1}$ J. A. Pechkis,$^{1,2}$ and P. L. Gould$^{1,*}$}
\address{$^1$Department of Physics, University of Connecticut, Storrs, CT 06269, USA
\\$^2$Currently with the Naval Research Laboratory,\\4555 Overlook Avenue S.W.,
Washington, DC 20375, USA\\$^*$Corresponding author: phillip.gould@uconn.edu}

\begin{abstract}We use a waveguide-based electro-optic phase modulator, driven by a nanosecond-timescale arbitrary waveform generator, to produce an optical spectrum with an arbitrary pattern of sidebands. A programmed sequence of linear voltage ramps, with various slopes, is applied to the modulator. The resulting phase ramps give rise to sidebands whose frequency offsets relative to the carrier are equal to the slopes of the corresponding linear phase ramps. This extension of the serrodyne technique provides multi-line spectra with sideband spacings in the 100 MHz range.
\end{abstract}


 ] 

\noindent
Control of a laser's frequency is an important capability for many applications in optical communications and atomic and molecular physics. It is often desirable to precisely shift the optical frequency or to expand a single optical frequency into several. An acousto-optic modulator can generate a frequency-shifted beam, but the efficiency falls off quickly above a few hundred MHz. Electro-optical phase modulators (EOMs), when driven sinusoidally, can provide multiple sidebands, with waveguide-based units allowing modulation at frequencies up into the tens of GHz. Sinusoidally modulating the injection current of a diode laser can also be used to generate sidebands \cite{Kowalski2001}. The optical spectrum resulting from sinusoidal modulation is a set of equally spaced sidebands with variable amplitudes based on the depth of modulation. Other methods of generating multiple frequencies have been demonstrated, but they rely on using individually tuned laser sources \cite{Ferrari1999}.  Frequency combs \cite{Ye2005} and pulsed serrodyne modulation \cite{Tomita2005, Tomita2003, Sanjoh2003} can provide a large number of frequencies at evenly spaced intervals from a single laser source, but these methods lack the ability to directly control the distribution of power.  Spectral shaping of high-bandwidth femtosecond pulsed lasers is easily accomplished with spatial light modulators and other techniques \cite{Weiner2000}, but slower timescales require alternative methods.

An interesting variation on electro-optical modulation is the serrodyne technique whereby a frequency shift is produced by driving a phase modulator (PM) with a sawtooth voltage \cite{Johnson1988}. Since frequency is the time derivative of phase, the slope of the linear phase variation gives the frequency offset. Recently, wideband single frequency shifting using serrodyne modulation has been demonstrated \cite{Johnson2010, Houtz2009}. Nonlinear transmission lines were used to produce the sawtooths, allowing frequency offsets in the GHz range. Here, we present a related method, extending the serrodyne technique to include a sequence of ramps with various slopes. This produces a pattern of sidebands whose offsets from the carrier are determined by the slopes of the corresponding phase ramps. The constraint of equal sideband spacings is thereby removed. Of course, there are limitations on the sideband widths and spacings from both Fourier considerations and the maximum phase change available with the EOM.
\begin{figure}[htb]
\centerline{
\includegraphics[width=8.2cm]{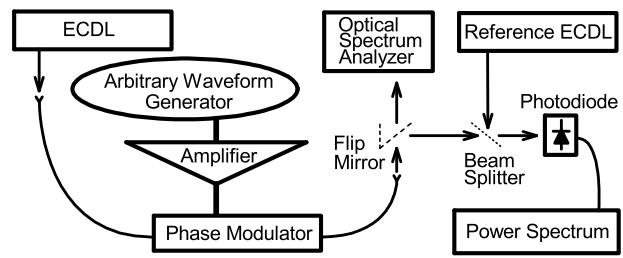}}
\caption{Experimental setup.}
\end{figure}

The experimental setup for generating and analyzing arbitrary spectra is outlined in Fig. 1. The main laser is a Hitachi HL7852G diode set up as an external-cavity diode laser (ECDL) \cite{Ricci1995}, yielding an output power of 30 mW tuned near 780 nm. A portion of the light from the ECDL is launched into a fiber and then coupled to the input pigtail of a lithium niobate waveguide-based electro-optic PM: EOSpace model PM-0K1-00-PFA-PFA-790-S. The PM is driven with a Textronix AFG3252B 240 MHz arbitrary waveform generator (AWG) whose output can optionally be amplified by a Minicircuits ZHL 1-2W amplifier with 500 MHz of bandwidth. The PM is internally terminated with 50 $\Omega$ and can handle an RF drive power of 30 dBm. Its phase response is characterized by the voltage necessary for a $\pi{}$ phase shift: $V_\pi{}$ $\simeq 2.0$ V (at 1 GHz). Our diagnostics include a 300 MHz free spectral range optical spectrum analyzer (OSA) with a linewidth of 1.5 MHz, as well as a heterodyne setup which can resolve the underlying structure of the sidebands. For the heterodyne analysis, the output of the PM is combined with a separate reference ECDL on a fast (2 GHz) photodiode. Although neither of the ECDLs is actively stabilized, their linewidths are narrow enough ($\sim$1 MHz) and the relative drift between them can be made small enough to yield sufficiently stable heterodyne signals over measurement times of 5 $\mu{}s$.  Labview programs utilizing the ``Auto Power Spectrum.vi" virtual instrument are used to Fourier analyze the measured heterodyne signals and to simulate power spectra from theoretical phase patterns.

For creation of arbitrary spectra consisting of several target frequencies, there are design considerations when building the phase pattern. First, the sawtooth ramp sequence corresponding to a particular frequency offset (from the carrier) $f_k$ should be calculated according to the prescription for serrodyne modulation. The phase ramps should have an amplitude of $2\pi{}n$ and a period of $1/f_k$ for a target frequency offset $nf_k$, where n is an integer. Second, due to Fourier constraints, the width of the envelope of a group of sidebands around a target frequency depends inversely on the amount of time spent on the corresponding ramps. Additionally, the smallest frequency spacings of the lines that make up the arbitrary spectra are fixed at the repetition frequency of the overall ramp sequence. As a simple example, Fig. 2(a) shows a phase pattern consisting of two segments, a single negative ramp and an equal duration of constant phase. The first segment, if repeated by itself, would correspond to a third order ($n=3$) serrodyne shift of -75 MHz.  The second segment corresponds to the carrier.  A simulation of the resulting power spectrum based on this phase pattern is shown in Fig. 2(b).  Since the repeat time of this phase pattern is 80 ns, each sideband lines up at some multiple of 1/80 ns = 12.5 MHz.  The FWHM of the group of sidebands around both the carrier and -75 MHz targets is roughly given by 1/40 ns = 25 MHz., i.e., the inverse of the time spent generating each offset. This width can be narrowed by including more consecutive ramps in the pattern, as shown below, with the tradeoff that the phase pattern repeat time will increase.

\begin{figure}[htb]
\centerline{
\includegraphics[width=8.2cm]{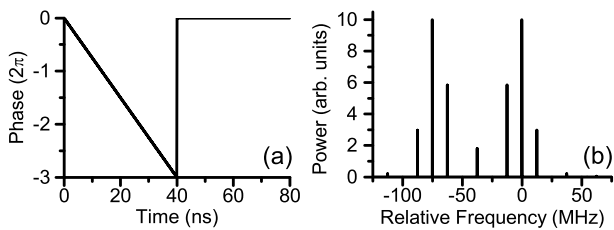}}
\caption{(a) One cycle of the sample phase pattern.  (b) Simulated spectrum based on (a).}
\end{figure}

To generate a spectrum with narrower widths for each target frequency (offsets of 0 and -75 MHz), a phase pattern similar to the one of Fig. 2(a) was constructed, but with the time allocated to each segment increased by a factor of six. The corresponding output of the AWG and the calculated power spectrum are shown in Figs. 3(a) and 3(b), respectively. With this phase pattern applied to the PM, The resulting spectrum measured with the OSA is shown in Fig. 3(c).  We see that the FWHMs of the target frequencies, in both the simulated and measured spectra, have been reduced to approximately 4 MHz, a factor of 6 narrower than in the simulated spectrum of Fig. 2(b), as expected. The tradeoff is in the time alternation between the target frequencies. In the limit of many consecutive identical ramps, only the frequency corresponding to those ramps is present during that time interval, and we recover the serrodyne condition. This is a fundamental limitation of this technique, imposed by Fourier considerations.
\begin{figure}[htb]
\centerline{
\includegraphics[width=8.2cm]{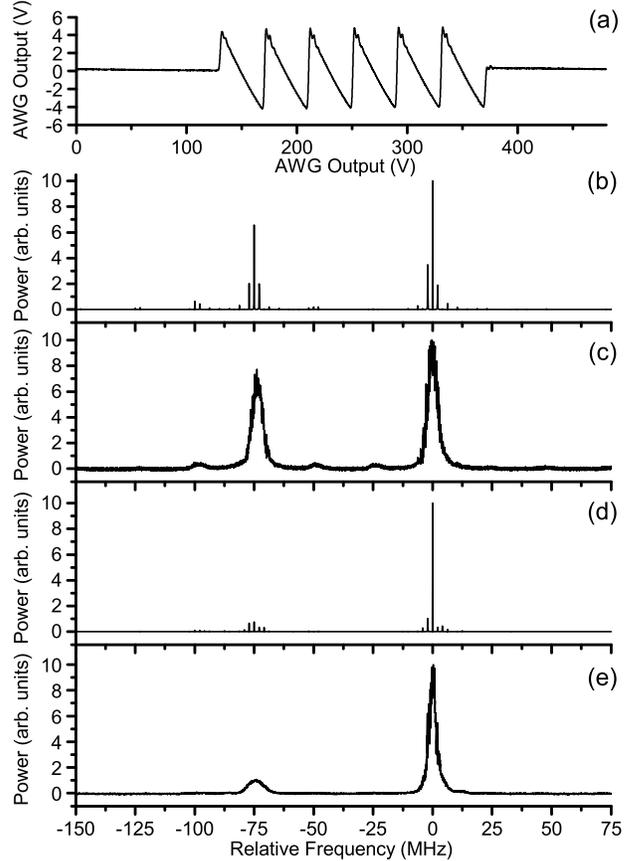}}
\caption{(a) One cycle of the AWG output of the phase pattern:  50\% ramps (-75 MHz),  50\% flat (carrier). (b) Simulated spectrum based on (a). (c) Measured spectrum with OSA.  (d) Simulated and (e) measured spectra from a phase pattern similar to (a) but with the duty cycle of the ramps reduced to 25\%.}
\end{figure} 

Because equal time is spent on the ramped and flat segments in Fig. 3(a), we expect equal powers at the two target frequencies: -75 MHz and 0 MHz (carrier). However, we see, in both the simulation and in the measurement, that the -75 MHz sideband contains less power than the carrier. We also see a small amount of power present at other frequencies (spurious sidebands). This infidelity is caused by imperfections in the sawtooth pattern due to the finite bandwidth of the AWG (and possibly impedance mismatch), as seen in the small ringing near the peaks. The 3 dB bandwidth of the PM is 15 GHz, so it does not contribute to the infidelity. If we simulate the spectrum from a perfect sawtooth (i.e., with zero reset time), it gives equal powers at -75 MHz and 0 MHz. These effects have been explored in depth for the case of pure serrodyne modulation \cite{Johnson1988}. Longer ramps with fewer resets would be desirable, but we are constrained by the maximum phase change of $6\pi{}$ achievable with our AWG and PM. We can control the relative power at the target frequencies by changing the durations of the ramps. In Fig. 3(d) (simulation) and Fig. 3(e) (measurement), we have modified the phase pattern of Fig. 3(a) so that the ramps occur over 25\% of the cycle, with the remaining 75\% of the cycle being flat. As expected, the power at -75 MHz offset is diminished. We also see that the peak widths are no longer equal, consistent with the Fourier considerations discussed above.
\begin{figure}[htb]
\centerline{
\includegraphics[width=8.2cm]{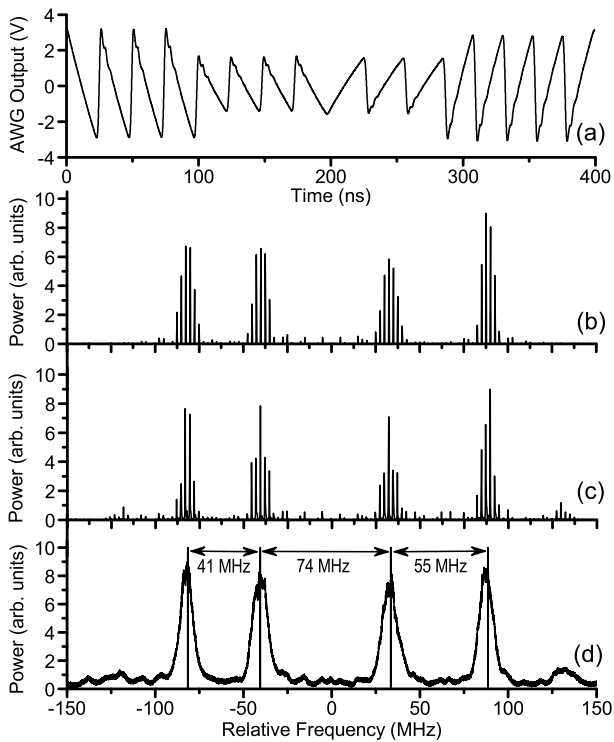}}
\caption{Example spectra of four target frequencies.  (a) Measured output of the AWG. (b)  Simulation of spectrum based on (a). (c) Spectrum from heterodyne measurement.  (d) Spectrum as measured with OSA. Vertical lines indicate target frequencies.}
\end{figure} 

Finally, we demonstrate how this modulation scheme can be used to generate multiple sidebands. As an example, we choose four target frequencies to match the hyperfine structure of SrF. Such light could be used to prevent optical pumping in the recently demonstrated laser cooling of these molecules \cite{Shuman2009, Shuman2010}. In the phase pattern, slopes to match the four target frequencies are incorporated and each segment is centered about zero to avoid distortion caused by the 5 MHz low-frequency cutoff of the RF amplifier. The programmed phase changes for the four ramps are: $4\pi{}$, $2\pi{}$, $2\pi{}$, and $4\pi{}$. The corresponding output of the AWG and the simulated spectrum are shown in Fig. 4(a) and Fig. 4(b), respectively. The resulting spectrum, measured with the heterodyne technique, is displayed in Fig. 4(c). The underlying sideband structure due to the 400 ns repeat time of the overall phase pattern is clearly visible. The spectrum measured with the OSA is shown in Fig. 4(d). The spacings between peaks nicely match those of the target frequencies. Note that the carrier has been completely suppressed. Since the phase pattern was designed to have approximately equal durations for each set of ramps, the four peaks are close to the same height.

In summary, we demonstrate a method of generating arbitrary spectra with an AWG and a fiber-based EOM. A spectrum containing four target frequencies unequally spaced over a 170 MHz span is presented as an example. A much broader frequency range and/or higher spectral fidelity should be achievable with a higher-bandwidth AWG. Since the generation of the phase pattern relies on an AWG, rapid tuning of the target frequencies is possible with an appropriately programmed waveform.  Although we use light at 780 nm, this technique could easily be adapted to wavelengths in the 700 - 2000 nm range, as fiber-based EOMs are commercially available at these wavelengths. This work was supported in part by the Chemical Sciences, Geosciences and Biosciences Division, Office of Basic Energy Sciences, U.S. Department of Energy.  We thank EOSpace for technical advice regarding the phase modulator.


\begin{thebibliography}{99}

\bibitem{Kowalski2001} R. Kowalski, S. Root, S. D. Gensemer and P. L. Gould, Rev. Sci. Inst. {\bf 72}, 2532 (2001).

\bibitem{Ferrari1999} G. Ferrari, M. Mewes, F. Schreck, and C. Salomon, Opt. Lett. {\bf 24}, 151 (1999).

\bibitem{Ye2005} J. Ye and S. T. Cundiff (eds.), \emph{Femtosecond Optical Frequency Comb: Principle, Operation and Applications} (Springer, 2005).

\bibitem{Tomita2005} I. Tomita, H. Sanjoh, E. Yamada, H. Suzuki and Y. Yoshikuni, J.  Opt. A {\bf 7}, 701 (2005).

\bibitem{Tomita2003} I. Tomita, H. Sanjoh, E. Yamada, and Y. Yoshikuni, IEEE Phot. Tech. Lett.  {\bf 15}, 1204 (2003).

\bibitem{Sanjoh2003} H. Sanjoh, I. Tomita, and Y. Yoshikuni,  Electron. Lett.  {\bf 39}, 392 (2003).

\bibitem{Weiner2000} A. M. Weiner, Rev. Sci. Inst.  {\bf 71}, 1929 (2000).

\bibitem{Johnson1988} L. M. Johnson and C. H. Cox, III, J. Lightwave Technol. {\bf 6}, 109 (1988).

\bibitem{Johnson2010} D. M. S. Johnson, J. M. Hogan, S.-w. Chiow, and M. A. Kasevich, Opt. Lett. {\bf 35}, 745 (2010).

\bibitem{Houtz2009} R. Houtz, C. Chan, and H. M\"{u}ller, Opt. Express. {\bf 17}, 19235 (2010).

\bibitem{Ricci1995} L. Ricci, M. Weidem\"{u}ller, T. Esslinger, A. Hemmerich, C. Zimmermann, V. Vuletic, W. K\"{o}nig, and T.W. H\"{a}nsch, Opt. Commun. {\bf 117}, 541 (1995).

\bibitem{Shuman2009} E. S. Shuman, J. F. Barry, D. R. Glenn, and D. DeMille, Phys. Rev. Lett. {\bf 103}, 223001 (2009).

\bibitem{Shuman2010} E. S. Shuman, J. F. Barry, and D. DeMille, Nature {\bf 467}, 820 (2010). 

\end{thebibliography}
\end{document}